# Fast and high-quality tetrahedral mesh generation from neuroanatomical scans

**Anh Phong Tran [1] and Qianqian Fang [2,*]**

[1]Department of Chemical Engineering and [2]Department of Bioengineering, Northeastern University, 360 Huntington Ave, Boston, MA, 02115, USA

**Abstract**. Creating tetrahedral meshes with anatomically accurate surfaces is critically important for a wide range of model-based neuroimaging modalities. However, computationally efficient brain meshing algorithms and software are greatly lacking. Here, we report a fully automated open-source software to rapidly create high-quality tetrahedral meshes from brain segmentations. Built upon various open-source meshing utilities, the proposed meshing workflow allows robust generation of complex head and brain mesh models from multi-label volumes, tissue probability maps, surface meshes and their combinations. The quality of the complex tissue boundaries is preserved through a surface-based approach, allowing fine-grained control over the sizes and quality of the mesh elements through explicit user-defined meshing criteria. The proposed meshing pipeline is highly versatile and compatible with many commonly used brain analysis tools, including SPM, FSL, FreeSurfer, and BrainSuite. With this mesh-generation pipeline, we demonstrate that one can generate 3D full-head meshes that combine scalp, skull, cerebrospinal fluid, gray matter, white matter, and air cavities with a typical processing time of less than 40 seconds. This approach can also incorporate highly detailed cortical and white matter surface meshes derived from FSL and FreeSurfer with tissue segmentation data. Finally, a high-quality brain atlas mesh library for different age groups, ranging from infants to elderlies, was built to demonstrate the robustness of the proposed workflow, as well as to serve as a common platform for simulation-based brain studies. Our open-source meshing software "brain2mesh" and the human brain atlas mesh library can be downloaded at http://mcx.space/brain2mesh.

## 1. Introduction.

The development of accurate and high-quality mesh models has taken a crucial role in today's volumetric medical image analyses. Three-dimensional (3D) mesh models are widely found in image visualization, image reconstructions from different modalities, and numerous finite element analyses (FEA) for physics-based modeling. The mesh representation possesses unique advantages over alternative approaches when discretizing a 3D domain. Although voxelated space has been the dominant form for processing a 3D imaging volume, some limitations are found when performing subsequent analyses. The terraced boundary shape in a voxelated space has difficulty in representing a smooth and curved boundary that delineates different tissues, resulting in a loss of accuracy. The uniform grid structure of the voxel space also demands a large number of cells in order to store anatomical structure without losing spatial details; this can cause prohibitive memory allocation and runtime in applications where solving sophisticated numerical models based on medical images is necessary. Another approach – octree – uses nested voxel refinement near curved boundaries. This improves memory efficiency significantly, but still suffers from terraced mesh boundaries (Frey and George, 2000; Owen, 1998; Thompson et al., 1999).

To resolve a complex structure using limited numbers of vertices and cells, many numerical techniques utilize a tetrahedral mesh model to gain computational and memory efficiency. This has led to the wide adoption of FEA where tetrahedral meshes are commonly used to effectively build linearized systems and approximate complex shapes with piecewise-linear basis functions. The feature-adaptive element size control in a tetrahedral mesh provides a mean to balance shape error and the size of the unknown space. In some cases, higher order elements such as hexahedral or curved elements can be employed, but they are usually more difficult to generate in a systematic manner for complex geometries. The gains in computing time by using these higher order elements are often offset by more sophisticated mesh generation processes.

## 1.1. Applications of mesh models in brain research and clinical care

Aside from extensive use in computer graphics and 3D rendering of brain structures, mesh models play key roles in modern neuroimaging modalities. In functional near-infrared spectroscopy (fNIRS), computationally extensive models of light propagation inside brain tissues must be solved quantitatively using the finite-difference (FD) (Klose and Hielscher, 2002), finite-volume (FV) (Ren et al., 2004), finite-element (Joshi et al., 2008) or mesh-based Monte Carlo (MMC) method (Fang, 2010) in order to recover brain activations. Similarly, electroencephalography (EEG) uses the conductivity properties of the head model and electrical potential measurements at a patient's head surface to monitor brain activities (Ramon et al., 2006). The effects of transcranial magnetic stimulation (TMS) and transcranial direct current stimulation (tDCS) can be simulated on realistic mesh models to evaluate brain damages (Yang et al., 2011) or by measuring its effects on major brain disorders (Rampersad et al., 2014). In addition, FE analyses of brain tissue deformation using mesh-models can assist neurosurgeons for operations or in the study of traumatic brain injuries and surgical planning (Warfield et al., 2002).

## 1.2. Segmentation of anatomical brain MRI scans

The first step towards generating a high-quality brain mesh model is to segment the brain into different tissues. While a manual segmentation was often performed, it is a lengthy process and can yield inconsistent results. With the improvements in template-based brain segmentation techniques, automated tools have become the preferred option in recent years. A list of neuroimaging software, such as FreeSurfer (Fischl, 2012), Statistical Parameter Mapping (SPM) (Ashburner and Friston, 2005), FMRIB Software Library (FSL) (Jenkinson et al., 2012), and BrainSuite (Shattuck and Leahy, 2002), can automatically segment the human brain from a T1-weighted MRI scan. A matching T2-weighted MRI can be used in some of these pathways to improve the segmentation accuracy.

It is important to note that the capabilities and focuses of these analysis tools vary greatly. FreeSurfer, for example, provides detailed cortical surfaces and parcellations of the subcortical structures, but does not provide a sulcal cerebrospinal fluid segmentation. In some instances, applications may only require the segmentation of the brain into white matter (WM), grey matter (GM) and cerebrospinal fluid (CSF) by skipping the computationally expensive step of calculating a detailed cortical surface model. In other instances, the WM and GM segmentations of the brainstem and cerebellum are not performed. Moreover, tools such as SPM and FSL can obtain additional segmentations such as scalp and skull that other tools may not provide. This suggests that segmenting a brain anatomy is often bounded by the purposes of the analysis. Therefore, it is beneficial and often necessary to have a flexible workflow that can combine outputs from various tools and be tailored to meet the needs of different research focuses (Perdue and Diamond, 2014).

One should also recognize a number of common challenges that arise when segmenting a brain MRI scan. For example, separating between the skull and CSF has been shown to be difficult in the absence of a T2-weighted MRI, as T1-weighted scans are mostly suited for differentiating the white and grey matters. Moreover, the default anatomical templates that are used in the brain segmentation software are often not appropriate for processing pediatric brains, in particular neonatal brains (Sanchez et al., 2012). Processing brains with lesions or containing diseases such as multiple sclerosis (De Stefano et al., 2014; Jain et al., 2015; Sormani et al., 2014) or Alzheimer's (Daianu et al., 2013) also remains an active area of research. Segmentations of brain tissues vary depending on the utilized neuroimaging tools; similarly, manual segmentations from experts also exhibit significant variations (Kazemi and Noorizadeh, 2014). In the absence of an accurate patient-specific brain segmentation, brain atlases are often used, including the Colin27 (Holmes et al., 2015) and ICBM-152 (Fonov et al., 2011) brain atlases or the UNC Infant brains (Shi et al., 2011). High-quality mesh models, especially pediatric brains, are not widely available due primarily to the lack of public atlas datasets until recently (Fillmore et al., 2015; Sanchez et al., 2012), and the lack of effective meshing tools. This has led us to believe that developing a high-quality tetrahedral mesh database for the brain would greatly facilitate some of the research analyses that do not require patient-specific models.

*1.3. Previous works on brain mesh generation*

The complexity of the brain anatomy and the algorithmic difficulties of generating high-quality mesh models limit the tools that are available for processing brain anatomical scans. These challenges become more evident when considering diverse criteria for measuring the quality of the mesh – such as the accuracy of the tissue boundaries, the quality of individual elements, and the density of the mesh. Many published methods have attempted to address part of the criteria, but have yielded suboptimal results in meeting other criteria. For example, a voxel conforming mesh generation approach (Lederman et al., 2011; Montenegro et al., 2009), the marching cubes algorithm (Lorensen and Cline, 1987; Wu and Sullivan, 2003) and "Cleaver" software (Bronson et al., 2013), can achieve good surface accuracy, but at the cost of highly dense elements near the boundaries due to the voxel-splitting or octree-like refinement. The large number of nodes generated from these approaches can cause excessive computational time when using the output meshes in FE analyses.

A general-purpose 3D mesh generation pipeline was proposed alongside with an open-source meshing software BioMesh3D (Callahan et al., 2009). This approach makes use of a point cloud and physics-based optimization to obtain a multi-material feature-preserving tetrahedral mesh through Delaunay tetrahedralization. This algorithm yields smooth boundaries with high-quality tetrahedral elements. The Delaunay-based meshing also allows fine-grained control on tetrahedral element sizes and yields more consistent element quality. While this approach can deal with complex volumes, the physics-based mesh optimization often results in lengthy run-time and may fail for the most difficult cases. In another Delaunay-based meshing pipeline provided in the Computational Geometry Algorithms Library (CGAL), the tetrahedral mesh is generated from a random point-set that is iteratively refined (Boltcheva et al., 2009; Pons et al., 2008)This procedure is relatively fast, parallelizable, robust for arbitrarily complex multi-label volumes, and provides controls on the tetrahedral element sizes and shapes, however, rough boundaries are often observed for meshing complex shapes. Obtaining detailed cortical folding structures is also challenging without having to generate an overly refined mesh. It also cannot take advantage of grayscale information, such as tissue probability maps, due to the label requirement

Commercially available tools, such as Mimics (Materialise, Leuven, Belgium), and ScanIP (Simpleware, Exeter, UK), offer integrated interfaces for image segmentation, mesh generation and manual mesh editing, but a streamlined high-quality mesh processing pipeline from neuroanatomical scans remains challenging. ScanIP was successfully applied to developing tetrahedral head mesh models from processed SPM segmentation outputs (Huang et al., 2013a). The segmentation and meshing pipeline proposed produced a mesh model containing approximately 7 million tetrahedral elements and a mesh generation time of 1-2 hours, even with the use of an adaptive meshing scheme to reduce the mesh density. This represents more than 1 tetrahedral element per $1\times1\times1$ mm$^3$ voxel. The highly dense mesh resulted in a similarly lengthy high-definition transcranial direct current stimulation (HD-tDCS). Through the Mimics software package, a combination of thresholding region growing and masking techniques can be used to obtain a segmented brain (Eggebrecht et al., 2012). A tetrahedral mesh can then be produced from this segmentation and further improved using their extension tool "3-matic". In addition to the licensing cost, this procedure requires extensive interactions with the software. Still, the built-in segmentation tools can yield suboptimal results when compared with more advanced template-based approaches used in dedicated brain analysis suites such as SPM and FSL.

In 2010, we developed a surface-based brain meshing approach and demonstrated that a high-quality multi-layered brain/head mesh can be built combining cortical and white matter surfaces from FreeSurfer and multi-label segmentations using our open-source meshing toolbox "Iso2Mesh" (Fang, 2010; Fang and Boas, 2009). In 2013, a similar approach was reported (Windhoff et al., 2013), in which an automated meshing pipeline "mri2mesh" was reported, incorporating FreeSurfer surface models and the scalp, skull, and CSF segmentations from FSL. A mesh-decoupling step built upon Meshfix (Attene, 2010) was introduced to avoid intersections between different tissue layers. With a non-intersecting multi-layered (i.e. nested) surface model assumption, mri2mesh permits a flexible surface-based meshing pathway, yielding smooth boundaries, high-quality tetrahedral elements, and control over the parameters of individual tissues. Even so, the reported meshing times are on the order of 3-4 hours, in addition to the cost of segmentation. Because mri2mesh is specifically developed using FreeSurfer and FSL outputs, it is also influenced by the inherent limitations of these tools, such as the inadequacy of bone segmentation using FSL Betsurf tool in the absence of a T2-weighted MRI scan.

*1.4. Proposed workflow*

In this work, we report "brain2mesh" – a fully automated, highly flexible, and mesh-quality-criteria driven brain/full head 3D meshing pipeline that has the advantages of being fast, robust, accurate, and simple-to-use. This work was built upon our well-disseminated MATLAB-based meshing toolbox, Iso2Mesh (Fang and Boas, 2009), with incorporations of additional open-source meshing utilities, such as CGAL (Boltcheva et al., 2009; Pons et al., 2008), Cork (Bernstein, 2017) and TetGen (Si, 2015). Our meshing toolbox accepts outputs from many common neuroimaging analysis tools and can handle various types of inputs including multi-label or probabilistic segmentations and surface models. Typically, it can create a high-quality tetrahedral mesh within a few minutes. The surface-based approach and the use of quality criteria permit fine-grained control on the mesh density. To demonstrate the ability to process a wide range of brain scans, we processed the recently published Neurodevelopmental MRI database (Fillmore et al., 2015; Sanchez et al., 2012) and created a library of tetrahedral mesh models of brain atlases with age ranging from newborns to elderlies. We also share this mesh library to the research community as an open-dataset, and strongly believe that it can facilitate the research in quantitative neuroimaging research, particularly on brain development.

2. **Material and methods**.
*2.1. Brain segmentation*

A diagram summarizing the common pathways in segmenting a neuroanatomical scan using popular neuroimaging analysis tools is shown in Fig. 1. In most cases, a tissue probability or multi-label volume is obtained for the WM, GM and CSF. Some software tools such as Statistical Parametric Mapping (SPM) allow the use of a matching T2-weighted MRI to improve the CSF segmentation. Utilities such as the FSL Brain extraction tool and SPM can provide additional information on the scalp and skull. In this work, we primarily focus on 6 tissue types – WM, GM, CSF, skull, scalp and air cavities. The flexibility of our approach is demonstrated using outputs from multiple popular neuroimaging tools. Additional classes of tissue (e.g. dura, vessels, fatty tissues, skin, muscle) are also available in some tools, which can be incorporated to our meshing processing steps. However, one should be aware that adding additional segmentations may result in increased node numbers and surface complexity, including disconnected surface components.

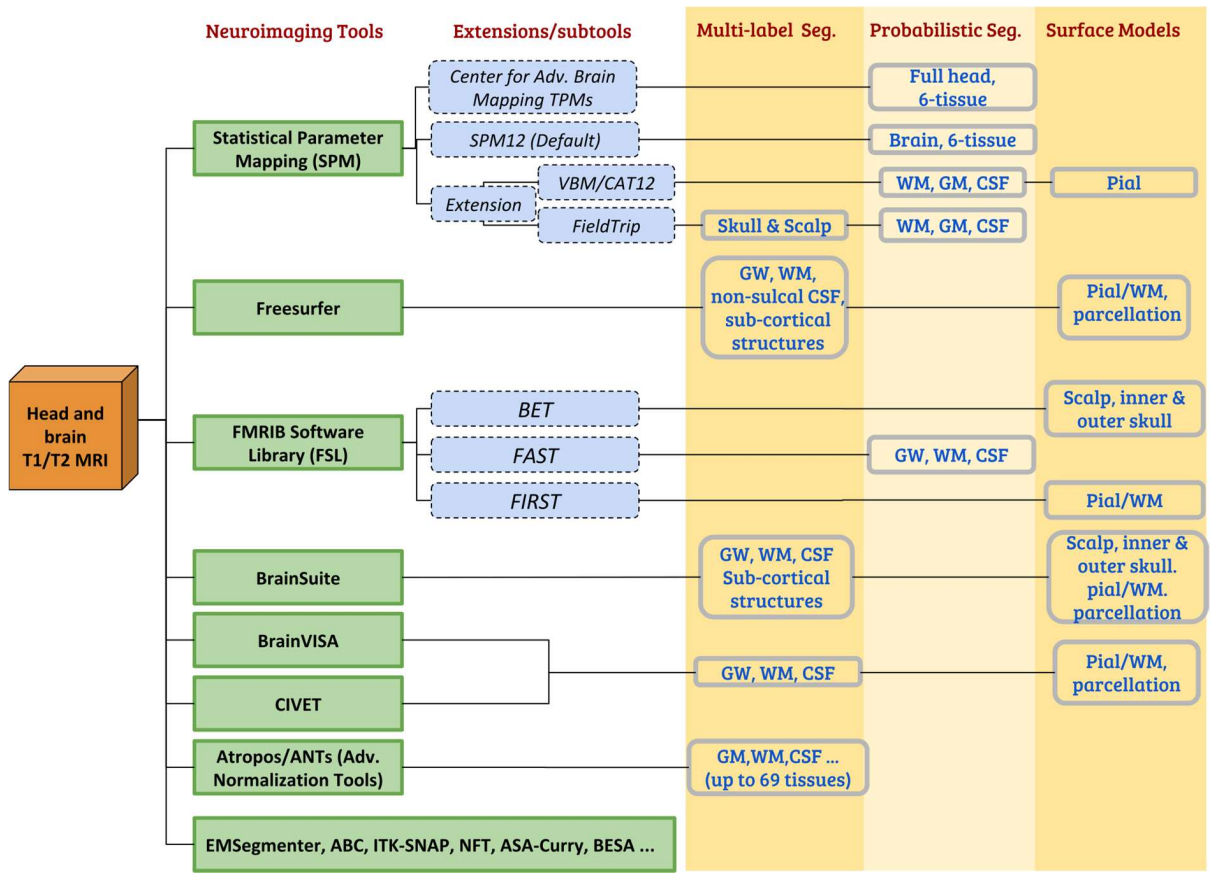

**Figure 1.** Segmentation pathways from anatomical head and brain MRI scans. The common neuroimaging tools/extensions (left) and the corresponding outputs (right, shaded) are listed.

*Segmentation pre-preprocessing*

The segmented brain or full-head volume, represented by either a multi-label mask or a set of probability maps (in floating-point values), are preprocessed to ensure a layered tissue model – i.e., the WM, GM, CSF, bone and scalp are incrementally enclosed by the later tissue layers, where the scalp surface is the outermost layer and the WM is the innermost layer. Such an assumption is also adopted in previous brain mesh generation literature (Burguet et al., 2004; Windhoff et al., 2013). We also consider air cavities, which are located between the outside skull and outside scalp surfaces. In Fig. 2, the workflow to create the different masks is outlined.

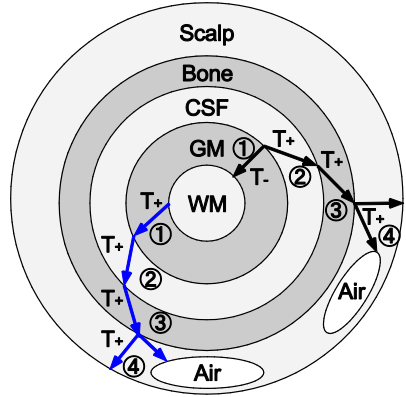

**Figure 2.** Illustration of the topological assumptions and the segmentation preprocessing workflow. Multiple air cavities are allowed. An arrow represents a thinning ($T_-$) or thickening ($T_+$) operation between two adjacent regions. Two sample pathways are indicated, shown by black and blue arrows, respectively. The circled numbers indicate the processing order.

In the regions where the above assumption is not satisfied, i.e. two adjacent tissue regions share a common boundary or the inner layer extends beyond the outer layer, the inner tissue probabilistic map (a 0-1 valued mask for binary segmentations) is truncated beyond the boundary of the outer tissue; in addition, a small gap is inserted between the two tissue boundaries to reinforce the above assumption. This is achieved using either a "thickening" or a "thinning" operator. In the case of a thickening operator ($T_+$), the outer layer tissue segmentation ($P_{out}$) is modified by

$$T_+(P_{out}; P_{in}): P_{out} \leftarrow \max [P_{in} + P_{out}, D_\varepsilon(P_{in})]$$

where $P_{in}$ and $P_{out}$ represent the inner and outer tissue probabilistic segmentation, respectively. $D$ denotes a "max-filter", i.e. a volume dilation operator defined by replacing each voxel with the maximum value in a cubic neighboring region with a half-edge length of $\varepsilon$, i.e.

$$D_\varepsilon(P): \quad P[i,j,k] \leftarrow \max(\{ P[i + \Delta i, j + \Delta j, k + \Delta k] \}_{-\varepsilon \leq \Delta i, \Delta j, \Delta k \leq \varepsilon})$$

Applying the $T_+$ operator defined above, a gap of width $\varepsilon$ is inserted between the inner and outer tissue layers. In this work, we set $\varepsilon=1$. To minimize the distortions to the original tissue segmentations, one can use a sub-voxel dilation operator, as defined by

$$D_\varepsilon(P; \alpha): P[i,j,k] \leftarrow \alpha P[i,j,k] + (1 - \alpha) \times \max(\{ P[i + \Delta i, j + \Delta j, k + \Delta k] \}_{-\varepsilon \leq \Delta i, \Delta j, \Delta k \leq \varepsilon})$$

where scalar $\alpha \in [0,1]$. The larger the $\alpha$ value, the less distortion to the original segmentation; however, if the two tissue boundaries are too close, it may cause intersections when combining the extracted surfaces in the later steps. Similarly, a "thinning" operator ($T_-$) is defined by

$$T_-(P_{in}; P_{out}): P_{in} \leftarrow \min [P_{in}, E_\varepsilon(P_{in} + P_{out})]$$

where $E_\varepsilon(P_{out})$ is an "erosion" operator of width $\varepsilon$, defined by a "min-filter" as

$$E_\varepsilon(P): \quad P[i,j,k] \leftarrow \min(\{ P[i+\Delta i, j+\Delta j, k+\Delta k] \}_{-\varepsilon \leq \Delta i, \Delta j, \Delta k \leq \varepsilon})$$

The $T_-$ effectively shrinks the inner layer mask at the intersecting regions with the outer layer. We note here that the above $T_+$ and $T_-$ operators work for both probabilistic segmentations and binary segmentations. We also highlight that the above operators only alter tissue segmentations in the areas where the inner tissue boundaries merge or intersect with the outer boundaries. Such regions only account for a small fraction of the brain tissue boundaries generated from realistic data.

### 2.2. Surface extraction and processing

If the input data is in the form of a multi-labeled volume or tissue probability maps (see Fig. 1), the next step of the mesh generation is to create a triangular surface mesh for each tissue layer. This is achieved using the "$\epsilon$-sample" algorithm using the CGAL Surface Mesh Generation library (Boissonnat and Oudot, 2005). For each extracted tissue surface, independent mesh quality and density criteria can be defined. In general, a surface mesh extracted from a probabilistic segmentation (grayscale) is smoother than one derived from a binary segmentation. When higher smoothness on extracted surfaces is desired, one can apply one of the 3 smoothing algorithms provided in Iso2Mesh, including the Laplacian, Laplacian+HC and low-pass filters (Bade et al., 2006). Although the pre-processing step described in the above section reduces the chance of intersection between adjacent layers, the coarse sampling of the surface may still result in occasional intersections. In such case, the intersections can be resolved using a modified Cork 3D surface Boolean library (Bernstein, 2017). The surface resolving step may create small enclosed regions between two intersecting surfaces. To remove these artifacts, we call TetGen to enumerate each enclosed region in the combined mesh, calculate an interior point of each region, and perform a "point-in-surface" test to identify the inner-most surface that encloses each region. All regions that belong to the same tissue layer are then merged. Furthermore, a bounding-box is computed and used to truncate the merged full-head surface mesh to produce a flat boundary near the neck region.

If the input data contain triangular surfaces of one or multiple tissue regions, additional surface processing is often required. For example, if the pial and white matter surfaces contain a separate surface for each brain hemisphere, a "merge" operation is needed to combine them into one closed surface. In another case, the ventricles (filled with CSF) are separated to the CSF region by a closed pial and WM surfaces generated by BrainSuite and FreeSurfer. The ventricle surface can be created by using the labeled volume map of the subcortical structures. To avoid intersection between the ventricles and pial/WM surfaces, the ventricle surfaces are pushed inward slightly to avoid intersection with the WM surface. This is achieved using the "decoupling" operator provided by Meshfix (Attene, 2010). Additionally, the FreeSurfer and FSL pial/WM surfaces do not include cerebellum and brainstem regions. To add those two anatomical regions, we first rasterize the pial/WM surfaces and then subtract those from the GW/WM probabilistic segmentations. From the subtracted probability maps, we then extract the brainstem and cerebellum white matter and grey matter surfaces. Like the ventricles surfaces, the cerebellum and brainstem surfaces are decoupled to the pial surfaces using Meshfix to ensure that there is no intersection.

### 2.3. Volumetric mesh generation and post-processing

With the derived combined multi-layer head surface model from above, a full head tetrahedral mesh can be finally generated using a constrained Delaunay tetrahedralization (CDT) algorithm, performed by calling an open-source meshing utility TetGen (Si, 2015). TetGen allows one to specify a number of meshing criteria to control the output mesh quality and density. An unnormalized radius-edge ratio ($q$) lower bound (see next section) can be specified to control the overall quality of the tetrahedral elements. In addition, one can set an upper-bound for the tetrahedral element volume globally for the entire mesh, or for a particular tissue label. An even more flexible approach is to define a "sizing field" by specifying the maximum element volume adjacent to every vertex in the input surface. After the tessellation, each enclosed compartment in a multi-layered input surface mesh is filled by tetrahedral elements, and is assigned with a unique integer label. One can manually specify the label for each enclosed region by a user-specified seed point, or let TetGen to enumerate all regions automatically.

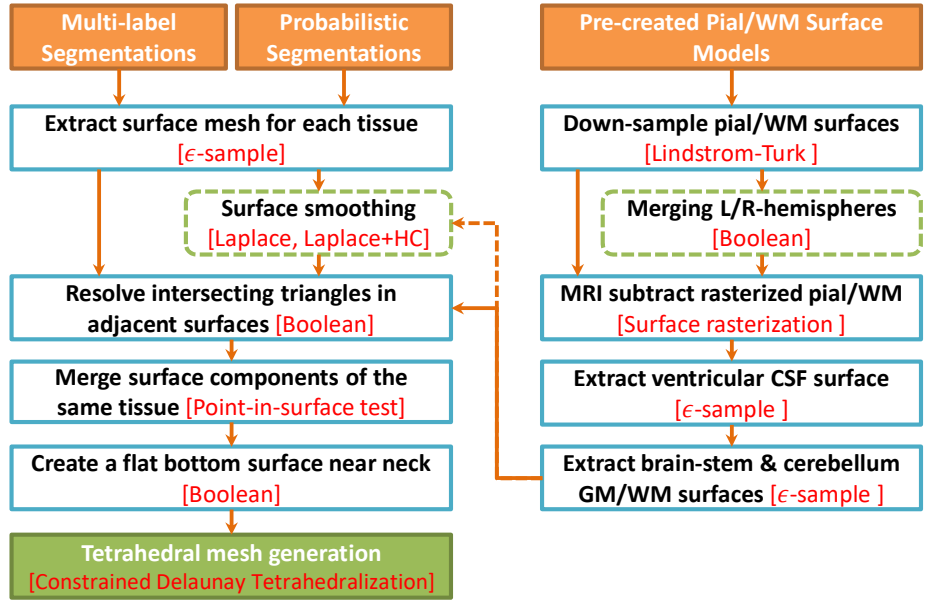

**Figure 3.** Processing steps for a surface-based mesh generation workflow. The left-side shows the steps for processing tissue probability maps and multi-label volumes, and the right-side shows additional steps to incorporate pre-created pial and white matter surfaces. Algorithms used in each step are shown in red.

2.4. *Mesh quality evaluation*

To assess the equality of the generated tetrahedral mesh, element shape metrics are used. The Joe-Liu quality metric (Liu and Joe, 1994) is defined as:

$$\eta = \frac{12 \times (3V)^{\frac{2}{3}}}{\sum_{0 \leq i < j < 3} l^2_{ij}}$$

where $V$ is the volume of a tetrahedron and $l_{ij}$ denotes the edge length between the $i$-th and the $j$-th vertices in the tetrahedron. Moreover, the normalized radius-edge ratio is defined as a ratio between a tetrahedron's inscribed sphere radius ($r_{in}$) and its longest edge ($l_{max}$) (Du, 2005):

$$Q = 2\sqrt{6}\left(\frac{r_{in}}{l_{max}}\right)$$

A third normalized quality metric $\rho$ is defined by the ratio between the radius of a tetrahedron's inscribed sphere radius ($r_{in}$) and that of its circumsphere ($r_c$):

$$\rho = 3\frac{r_{in}}{r_c}$$

All above 3 metrics range between 0 and 1, with 1 being achieved at the highest quality (equilateral tetrahedron) and 0 at the worst-case scenarios (i.e. degenerated element). We want to note that an un-normalized "*radius-edge ratio*" is used by TetGen as the refinement termination condition (Si, 2015). It is defined by the ratio between the tetrahedron's circumscribed sphere radius ($r_{out}$) and its shortest edge ($l_{min}$).

$$q = \frac{r_{out}}{l_{min}}$$

This unnormalized radius-edge ratio has a theoretical lower-bound at 0.612 (at an equilateral tetrahedron). Opposite to the other three normalized metrics, a small $q$-value indicates well-shaped elements.

3. **Results and Discussion**

In the below subsections, we showcase the robustness and the flexibility of our brain meshing pipeline described above by processing a range of complex brain anatomical scans. We first demonstrate various meshing pathways of our meshing pipeline by using selected datasets, and then report a high quality brain atlas mesh library ranging from newborns to seniors derived from the Neurodevelopmental MRI database (Fillmore et al., 2015; Sanchez et al., 2012). Finally, we show that our meshing algorithm is robust and can be used to process non-human animal brain MRIs. All computational times were benchmarked on an Intel i7-6700K processor using a single thread.

*3.1. High quality tetrahedral meshes of human head and brain models*

By applying the above meshing pipeline, we have created a set of high quality brain and full head meshes from our benchmark datasets. In Fig. 4, a mesh is generated from an SPM segmentation performed using tissue priors from the Laboratory for Research in Neuroimaging (LREN) (Lorio et al., 2016). The generated mesh contains 76,688 nodes and 462,646 tetrahedral elements, with average mesh quality metrics $\eta = 0.7454 \pm 0.1405$, $Q = 0.7001 \pm 0.1637$, and $\rho = 0.6021 \pm 0.1452$. A 1×1×1 mm3 T1-weighted MRI scan of an average head for the 40-44 years old age group from the University of South Carolina (USC) Neurodevelopmental MRI database (in the following sections, we will use "USC age-bracket" to refer to an atlas in this database. For example, the atlas used in this example is USC 40-44) was used as the original image. The segmentation yields 6 tissue classes that are used in the mesh generation: WM, GM, CSF, bone, scalp and air cavity.

The surface mesh extraction is performed with tissue-layer-specific maximum Delaunay sphere radii (Boissonnat and Oudot, 2005): 2 mm for the pial and white matter surfaces, 3 mm for the CSF layer, and 3.5 mm for the other tissue layers. The extraction, meshing and labeling steps took 26.30 sec of running time. The maximum element volume size was defined as 40 mm3, the radius-to-edge ratio was 1.414, and a mesh sizing field value of 4 is used to coarsify the mesh at regions where two surfaces almost overlap. The latter value is an option that can be used in TetGen to simplify overly dense areas in the final mesh. The probability threshold for surface extraction was set at 0.5 for each of the tissues. The meshing algorithm is implemented with single-threaded computing; therefore, multiple instances of mesh generation can be executed in parallel to process larger databases.

The tissue surfaces shown in Fig. 4 are generally smooth and of good quality. This is confirmed by the mesh quality metric histograms in Figs. 4(g-i). The mean values are close to 1 and no degenerated elements are found. Although it is possible to further improve the mesh quality through refinement, one should be cautious that a high density tetrahedral mesh can result in slow computation in any subsequent modeling tasks. The element volumes in Fig. 4(j) are well distributed with only a small portion of relatively small elements.

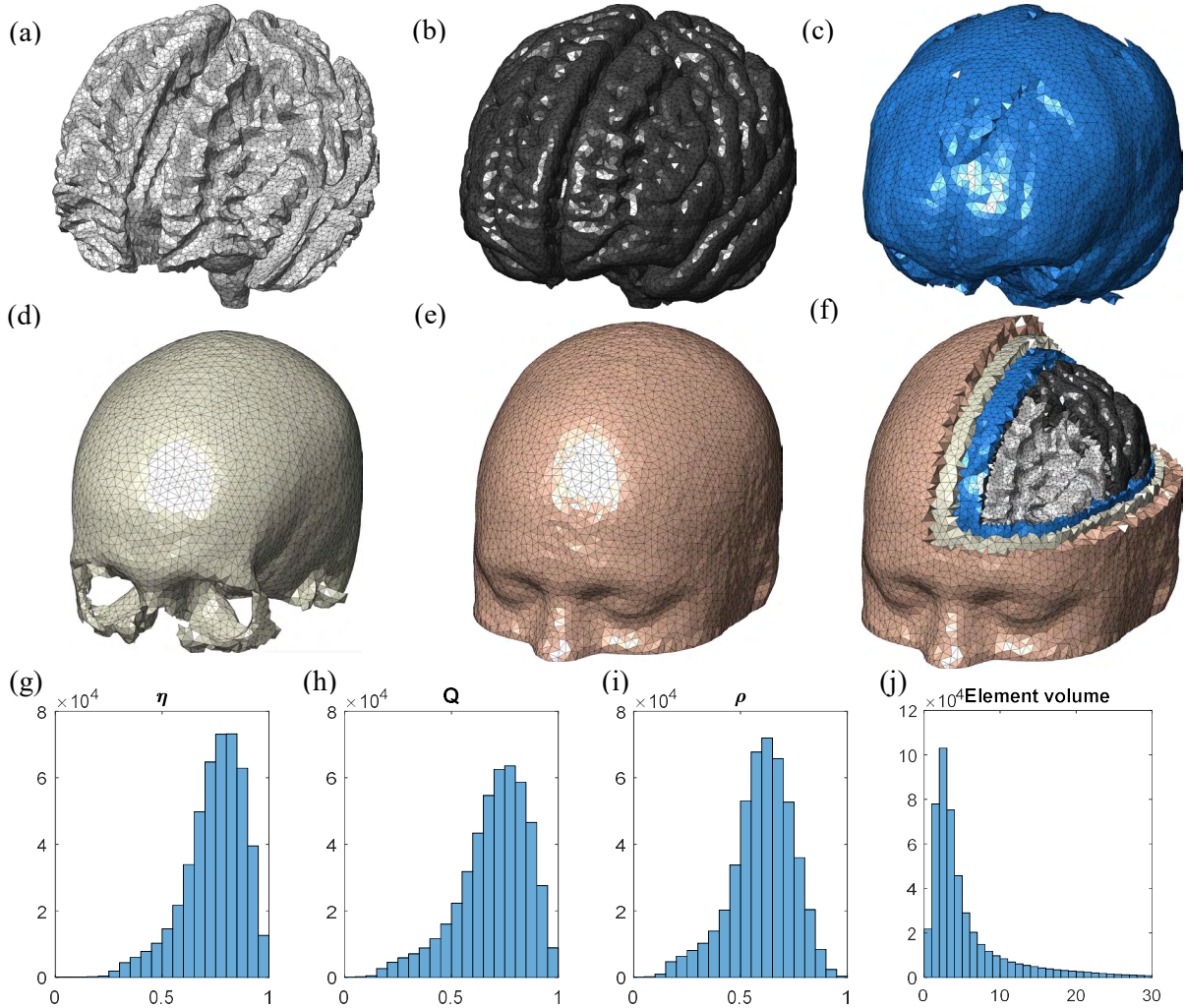

**Figure 4.** A 5-layer full head tetrahedral mesh derived from an atlas head of the USC 40-44 atlas. It contains (a) WM, (b) GM, (c) CSF, (d) bone, and (e) scalp layers. A cross-cut view of the tetrahedral mesh is shown in (f). We also show the histograms of the (g) Joe-Liu metric ($\eta$), (h) normalized radius-edge ratio ($Q$), (i) inscribed-to-circumscribed sphere radius ratio ($\rho$) and (j) element volumes.

In Fig. 5, we show a full-head mesh derived from the SPM segmentation of a 19.5 years old average head MRI scan provided by the Center for Advanced Brain Imaging (CABI) (Huang et al., 2013a). The final mesh contains 539,330 tetrahedral elements and 89,721 nodes. The parameters and tissue classes used are the same as the previous example. The quality metrics for the tetrahedral elements are $\eta = 0.7437 \pm 0.1481$, $Q = 0.6976 \pm 0.1710$, and $\rho = 0.6008 \pm 0.1500$. The algorithm took 26.02 seconds to create this whole-head mesh.

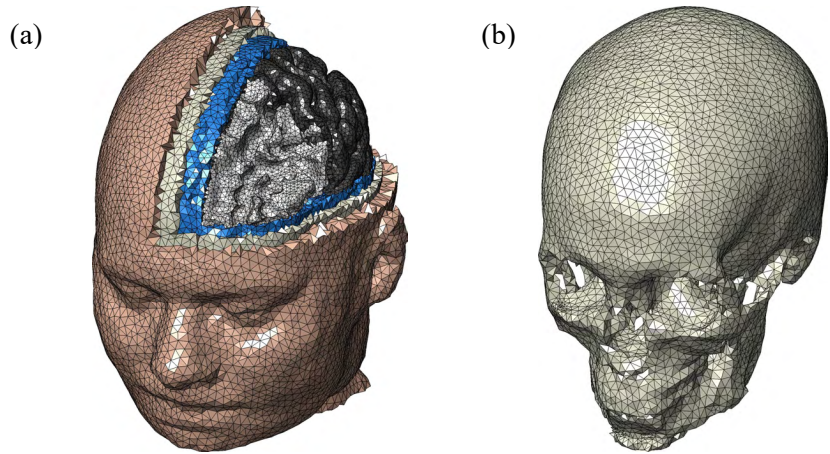

**Figure 5.** Tetrahedral mesh generated from a whole head segmentation using SPM. The segmentation data is derived from the CABI 19.5-year-old atlas. In (b), the surface mesh of the skull is shown.

To demonstrate that our meshing pipeline can interoperate with multiple segmentation tools, the head mesh in Fig. 5 is recreated using FSL segmentations in Fig. 6. In this case, the FSL Betsurf utility failed to produce a valid skull surface, therefore only the WM, GM, and CSF segmentations are used (Jenkinson et al., 2012). In addition, the scalp layer is extracted using Iso2Mesh from the raw MRI scan. The final mesh in Fig. 6 contains 462,868 tetrahedral elements, 76,378 nodes, with quality metrics of $\eta = 0.7577 \pm 0.1393$, $Q = 0.7127 \pm 0.1637$, and $\rho = 0.6131 \pm 0.1460$. The meshing process took 23.35 seconds. The mesh quality and density values are similar to the meshes produced using the SPM segmentation with LREN priors used in Fig. 4.

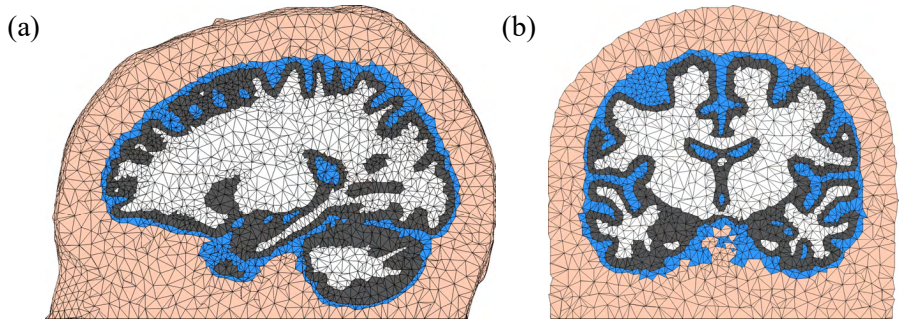

**Figure 6.** Tetrahedral mesh generated using FSL segmentation of the CABI 19.5-year-old atlas. The (a) sagittal (a) and (b) coronal (b) views are shown.

As mentioned above, our meshing pipeline can also process multi-label volumes (i.e. piecewise-constant tissue segmentations). In Fig. 7, we show a full head mesh generated from the segmented Colin27 atlas (Holmes et al., 2015). The segmented head volume has a $0.5 \times 0.5 \times 0.5$ mm$^3$ voxel resolution, with 6 tissue types. To apply our workflow, the dura and CSF volumes were merged together while the marrow and skull tissue classes were combined into the bone layer and the fat, muscle, skin-muscles labels were merged into the scalp layer. Using similar meshing criteria as in the previous examples, the final tetrahedral mesh was produced, containing 874,421 elements and 144,547 nodes with quality metrics of $\eta = 0.7163 \pm 0.1613$, $Q = 0.6646 \pm 0.1862$, and $\rho = 0.5746 \pm 0.1597$. Despite the significant increase in voxel density compared to the examples in Figs. 4-6, the total processing time was only 85.77 seconds.

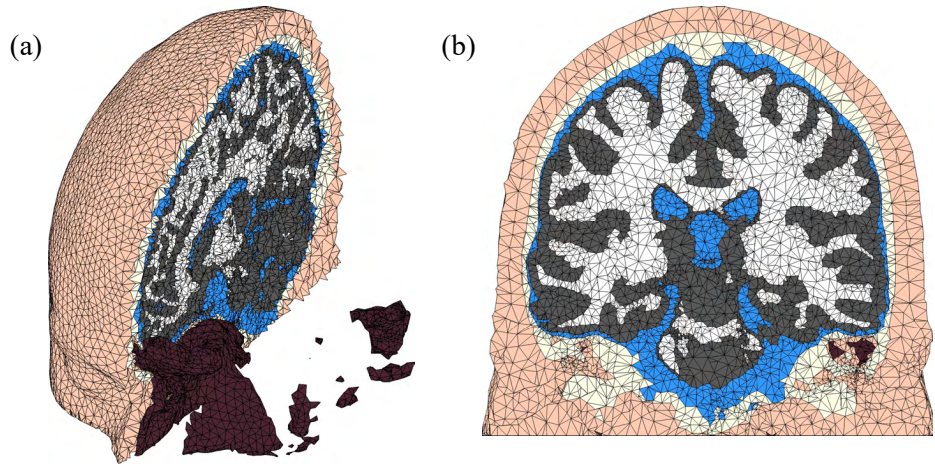

**Figure 7.** Tetrahedral mesh generated from the segmented (multi-label) Colin27 atlas. The (a) sagittal and (b) coronal views are shown.

To show the generality of our meshing workflow, in Fig. 8 (a), the WM, GM, and CSF segmentations of the "Haiko 89" baboon brain (Love et al., 2016) were used to produce a tetrahedral mesh model. The output mesh contains 229,388 elements, 38,754 nodes, and mesh quality metrics of $\eta = 0.7388 \pm 0.1468$, $Q = 0.6916 \pm 0.1682$, and $\rho = 0.5955 \pm 0.1477$. The meshing only took 11.76 seconds. In Fig. 8 (b), a similar segmentation for an ovine brain was processed to generate a tetrahedral mesh of 238,020 tetrahedral elements, 40,510 nodes, with $\eta = 0.7446 \pm 0.1385$, $Q = 0.6972 \pm 0.1625$, and $\rho = 0.6001 \pm 0.1438$. The processing time was 16.54 seconds. The 0.5 mm resolution segmented ovine brain from the McConnell Imaging Brain Imaging Centre (Nitzsche et al., 2015) was used here

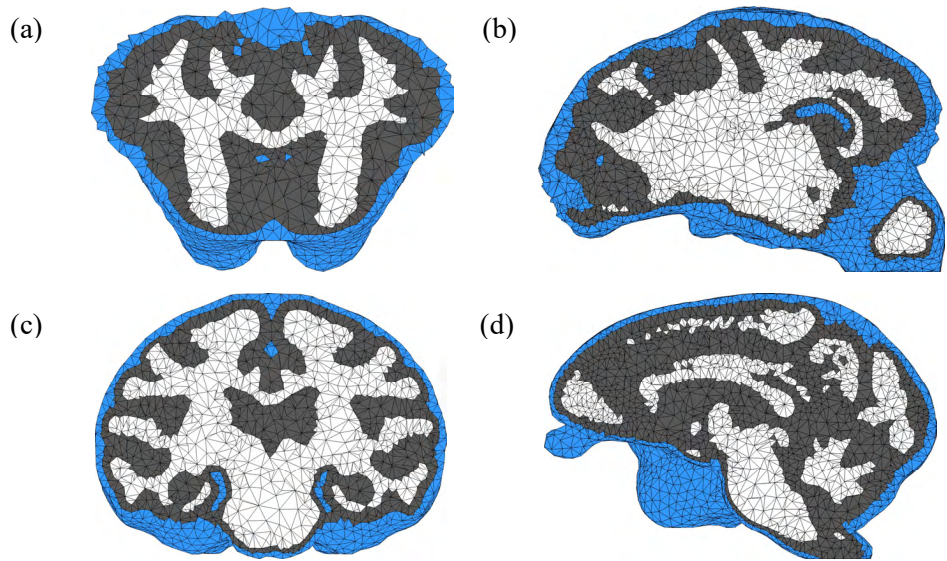

**Figure 8.** Mesh generated from the "Haiko 89" baboon dataset with (a) a coronal view and (b) a sagittal view. We also show the mesh generated from the McConnell Brain Imaging Center "Ovine Brain Atlas" with (c) a coronal view and (d) a sagittal view.

*3.2. Hybrid meshing pipeline combining volumetric segmentations with tissue surface models*

In this subsection, we demonstrate our "hybrid" meshing pathway. This approach can combine tissue surfaces extracted from probabilistic segmentations with the pial and WM surfaces generated by other neuroanatomical analysis tools, such as FreeSurfer and FSL. In the case of FreeSurfer, the pial and white matter surfaces provided can be very dense. In comparison, the pial surfaces we produced in the FSL and SPM examples in Figs. 4-6 are significantly coarser. To avoid creating extremely dense tetrahedral mesh and long mesh generation time, a mesh simplification process using the Lindstrom-Turk algorithm is used (Dey et al., 1999; Garland and Heckbert, 1997). In Fig. 9, we characterize the tradeoff between mesh size and the surface error calculated by down-sampling the FreeSurfer-generated pial surface for the USC 30-34 atlas using an open-source program "Metro" (Garland and Heckbert, 1997). At resampling ratio values below 0.1 (i.e. decimating over 90% edges), both gyri and sulci show a surface error above 0.5 mm. When the resampling ratio value is increased to 0.15, the observed error at the gyri becomes minimal, but few areas at sulci show errors above 1 mm. A resampling ratio of 0.2 is selected in this example provides typical gyri errors below 0.2 mm and only a handful of small sulci regions are shown to have errors above 1 mm.

As we illustrate in Fig. 3, a number of additional processing step have been taken to process the FreeSurfer pial/WM surfaces. These include the merging of the left and right hemisphere surfaces, addition of the CSF ventricles, and the addition of cerebellum and brainstem using SPM segmentations. The final mesh contains 917,212 elements, 150,999 nodes with $\eta=0.6808 \pm 0.1761$, $Q = 0.6340 \pm 0.1957$, and $\rho=0.5465 \pm 0.1670$. The mesh generation time was 162.42 seconds. The majority of the runtime was spent on repairing, decoupling and combining the FreeSurfer pial and white matter surfaces. This meshing time is significantly lower than the reported 3-4 hours required for creating a similar mesh using "mri2mesh" (Windhoff et al., 2013). While not shown here, this hybrid workflow also accepts probabilistic segmentations produced by FSL or pial/WM surfaces created by BrainSuite and similar combinations. Post-processing steps can be added using the labeled volume masks to reconstruct the corpus callosum between the two hemispheres and to bridge the brainstem to the cerebrum.

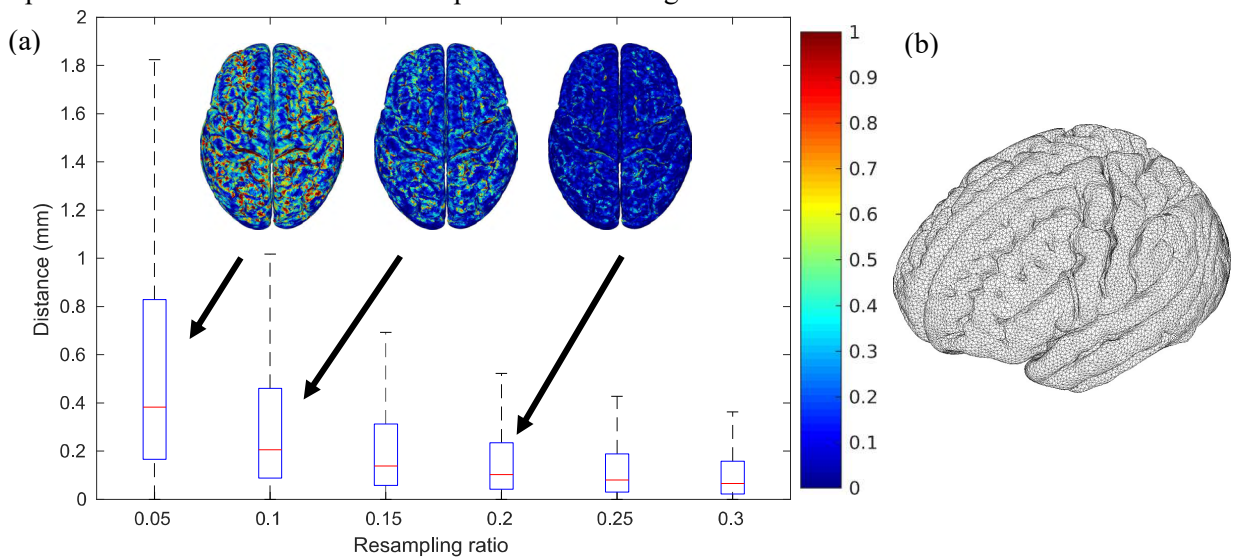

**Figure 9.** (a) Box-plot of surface errors as a function of resampling ratio (percentage of edges that are preserved) when down-sampling a FreeSurfer-generated pial surface. Spatial distributions of the errors are shown as insets. (b) A down-sampled pial surface mesh at a resampling ratio of 0.2.

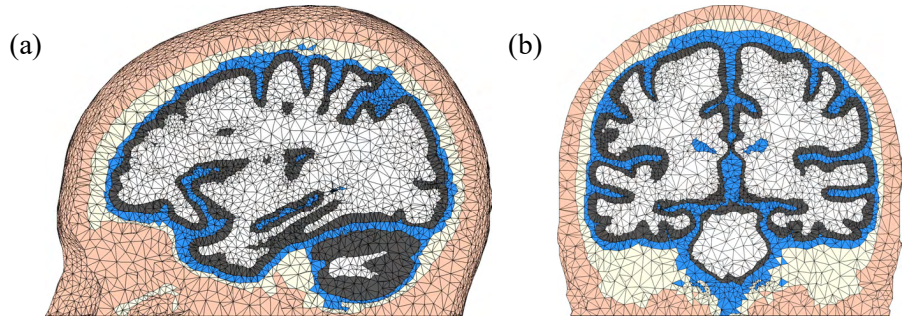

**Figure 10.** Tetrahedral mesh generated from a hybrid meshing pathway combining FreeSurfer surfaces with SPM segmentation outputs for the USC 30-34 atlas. The (a) sagittal and (b) coronal views are shown.

*3.3. Volumetric parcellation of the cortical region*

With a flexible mesh element mapping/labeling algorithm, a tetrahedral brain mesh can be easily partitioned into non-overlapping regions, i.e. "parcellations", according to different criteria. In Fig. 11, we show that the parcellation information derived from the BrainSuite output and detailed pial/WM surfaces can be combined to create a parcellated tetrahedral mesh. In the case presented here, segmentations from BrainSuite were created for the USC 30-34 brain atlas. After the mesh generation from the calculated surfaces, the tetrahedral elements are mapped to the BrainSuite parcellation volume and labeled with the according region index. The hybrid meshing pathway combining detailed white matter and cortical surfaces can be combined with this parcellation to give a full head mesh model with subcortical structures.

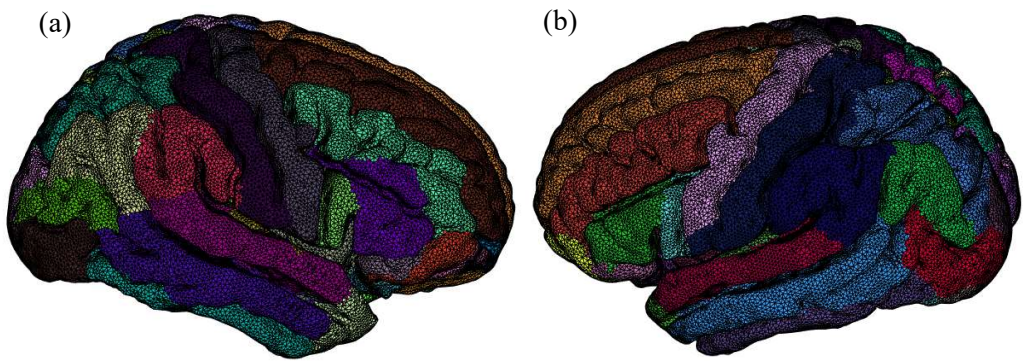

**Figure 11.** Parcellated tetrahedral mesh of (a) the right hemisphere and (b) the left hemisphere of the USC 30-34 brain atlas. The cortical surfaces and parcellation data were derived using BrainSuite.

*3.4. Library of high-quality tetrahedral meshes for developmental brain atlases*

Using our meshing workflow described above, we have processed the USC Neurodevelopmental MRI Database and generated full-head/brain meshes for all included age groups, ranging from newborns to 84-year-old. WM and GM segmentations provided by the database were used in all the meshes provided. For all atlases above 35-39 years-old, SPM was used to provide CSF and bone segmentations due to lack of availability in the USC database. When the scalp segmentation was unavailable, the scalp surface was extracted from the raw MRI data using a simple intensity thresholding approach followed by applying 3 iterations of Laplacian+HC smoothing algorithm (Bade et al., 2006). In Fig. 12, we show 15 sample USC atlas brain meshes as examples. In addition, we have also successfully processed the atlas database from BrainWeb (Aubert-Broche et al., 2006) (not shown). In all processed MRI scans, regardless of age, our meshing workflow worked smoothly; the average processing time is less than a minute per mesh when the voxel resolution is 1×1×1 mm$^3$ and about 3 min per mesh when the resolution is 0.5×0.5×0.5 mm$^3$.

In Fig. 13, we show the surface area and volume for WM, GM, and CSF calculated from the generated atlas meshes. The GM volume follows the reported inverted U-shaped pattern that peaks very early in the adolescence and the WM follows an increase in volume until the early twenties (Lenroot and Giedd, 2006). We show that from a mesh it is not only simple to deduce a volume of a tissue type, but that surface area calculations are greatly facilitated. While we did not specifically separate inner and outer surfaces of different tissues, nor did we calculate the surface area of a subset of these tissues, we believe that these are tasks that are more trivial to deal with using tetrahedral meshes.

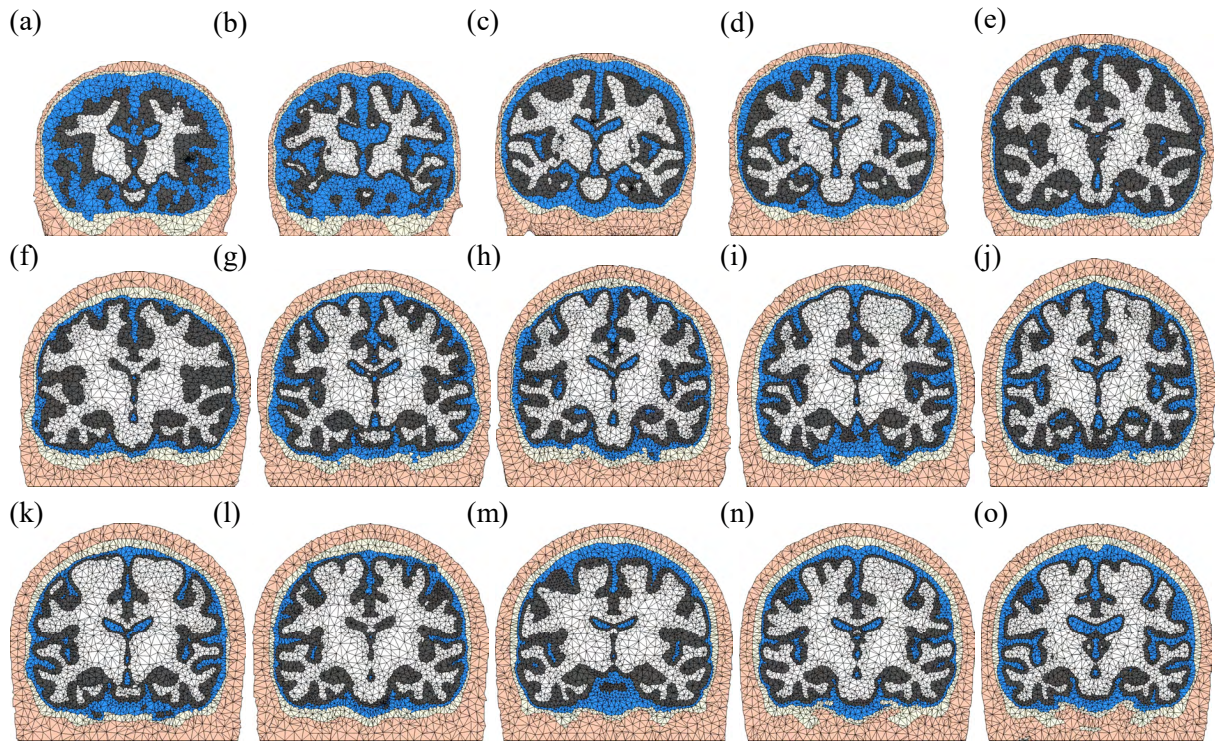

**Figure 12.** Tetrahedral brain meshes (coronal views) produced using the Neurodevelopmental MRI Database, including (a) 3 months, (b) 7.5 months, (c) 15 months, (d) 3 years, (e) 5 years, (f) 8.5 years, (g) 12 years, (h) 14 years, (i) 16 years, (j) 18 years, (k) 20-24 years, (l) 30-34 years, (m) 40-44 years, (n) 55-59 years, (o) 70-74 years.

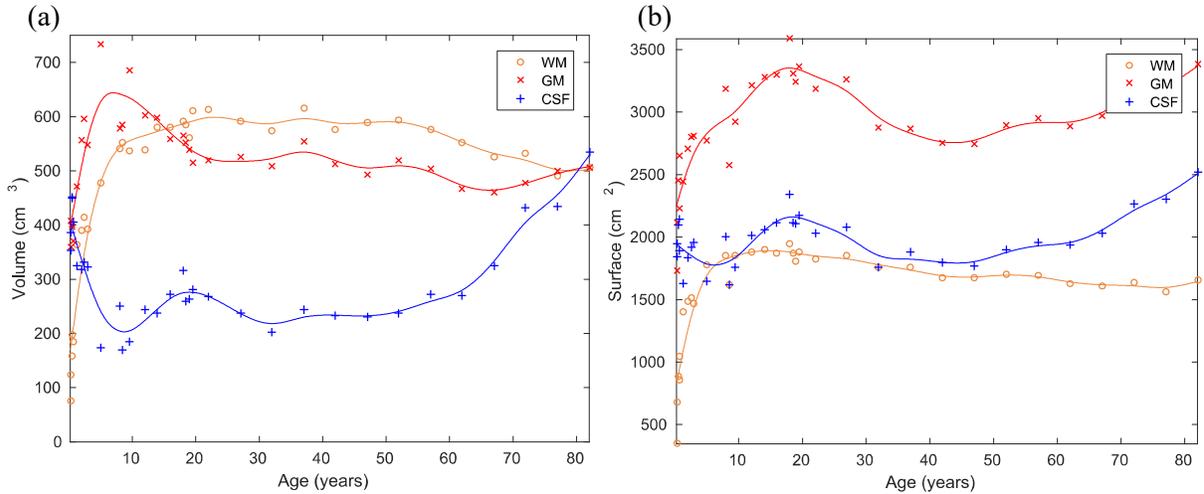

**Figure 13.** Effect of age on (a) the WM, GM and CSF volumes and (b) surface areas. The volumes and surfaces are only representative of the brain regions located inside the generated tetrahedral meshes. The smoothing splines (solid lines) are interpolated using a weighted coefficient of 0.05.

4. **Conclusion**

This work addresses the increasing needs for accurate and high-quality tetrahedral head mesh models that arise in many neuroimaging research disciplines for brain quantification, modeling, image reconstruction and visualization. Due to the intrinsic complexity of the brain anatomy and the diversity of neuroanatomical analysis tools, a meshing pipeline that is computationally efficient yet flexible enough to accommodate a wide range of segmentation inputs is of great value to the neuroimaging community.

In this paper, we described a fast and robust brain mesh generation algorithm and demonstrated that our MATLAB-based implementation of this algorithm, brain2mesh, can produce high quality brain and full head tetrahedral meshes from multi-label or probabilistic segmentations with full automation. The ability to create tissue boundaries from gray-scale probabilistic maps and incorporate detailed surface models from FreeSurfer/FSL ensures smoothness and high accuracy in representing brain anatomical structures, while necessary modifications to the tissue boundaries to satisfy our tissue topological assumptions were kept minimal and local. The output meshes generally exhibit excellent shape quality without needing to generate excessive number of small elements, which was a challenge often encountered in previously published meshing tools. This makes the output mesh computationally and memory efficient in subsequent data processing and modeling tasks. For most of the included examples, the processing time is less than 1 minute using only a single CPU thread. This is dramatically faster most other published brain meshing tools (Callahan et al., 2009; Huang et al., 2013b; Windhoff et al., 2013). Moreover, the entire meshing pipeline was developed based on open-source meshing utilities, including Iso2Mesh, CGAL, TetGen and Cork. This ensures excellent accessibility of this tool to the community.

The element size and quality of the output mesh are fully adjustable by tuning a group of explicit meshing criteria at various processing steps, including surface extraction, surface resampling, repairing, merging, and tetrahedral mesh generation. This allows users to generate meshes of different densities. Advanced meshing features, such as brain-parcellation based mesh partitioning, mesh refinement for specified labels or according to spatial functions (i.e. sizing fields) (Si, 2015) are also supported. With our optimized default meshing parameters, the entire process does not require interactive editing by the operator, and can be executed in a fully streamlined fashion.

Finally, we have successfully used this meshing tool to process many publicly available brain MRI datasets, including the USC Neurodevelopmental MRI database, BrainWeb atlases, and MNC atlases. In most cases, the meshing software has shown great robustness, and generated meshes of consistently high quality with a low run-time. The estimated GM/WM volumes derived from the generated meshes show a reasonable correlation with the subject age. In addition to sharing our meshing software, we will make the generated brain atlas mesh models available to the community. We strongly believe that this brain atlas mesh library will facilitate studies of model-based neuroimaging modalities, and help address important questions related to brain development with quantitative imaging analysis approaches. Our open-source meshing software and the human brain atlas mesh library can be downloaded at http://mcx.space/brain2mesh.

## 5. Acknowledgement

The authors wish to thank the funding supports from the National Institutes of Health [grant numbers R01-GM114365 and R01-CA0204443]. We acknowledge the valuable advice from Dr. Hang Si on the use of TetGen, as well as the instructive conversation with Dr. John Richards and Dr. Katherine Perdue on brain segmentation.